\documentclass[twocolumn,10pt]{asme2ej}

\usepackage{epsfig}
\usepackage{float}
\usepackage{amsmath}
\usepackage{gensymb}

\title{Predictive Control Based on Reduced Order Model for temperature homogeneity in a resin transfer molding tool for thermoset materials}

\author{Miguel Escolano \\
	{\tensfb Jos{\'e} Manuel Rodr{\'\i}guez} \\
	{\tensfb Javier Or{\'u}s} \\
	{\tensfb Manuel Laspalas} \\
	{\tensfb Agust{\'\i}n Chiminelli}
	\affiliation{
		ITAINNOVA \\
		Zaragoza, 50018 \\
		Spain \\
		mescolano@itainnova.es
	}
}

\begin{document}

\maketitle    

\begin{abstract}

{\it Resin Transfer Molding (RTM), which has attracted much attention in the last years for lightweight manufacturing, represents an important challenge in terms of control technology. During the process, a resin fills the cavity where a reinforcement fabric has previously been layered. This resin undergoes a chemical reaction which is thermically activated. Therefore, assuring a proper reaction requires a precise control of temperature in the entire mold cavity. Three factors make this control problem especially hard: the coupling among the large number of actuators and sensors, the variability of the test conditions and the power limitations of the electric actuators which do not offer cooling capability. The present work describes an optimized Model Predictive Control (MPC) architecture capable of handling these difficulties and also achieving the tight control requirements needed in the application. The thermal distribution inside the mold cavity is included into the controller by a simplified Reduced Order Model (ROM). This representation is obtained by data from an experimentally validated Finite Element Model (FEM), using AutoRegressive model with eXogenous terms (ARX) identification. In order to maintain the simplicity of this linear representation, the time-varying model parameters are estimated by using a perturbation observer. Additionally, the performance of the basic algorithm is improved: firstly, an augmented observer to estimate the temperature distribution of an extended spatial resolution; and secondly, a symmetry condition in the calculation of the control commands. The developed architectures have successfully been implemented in a RTM tool with the fulfillment of the control requirements.
}
\end{abstract}



\section{Introduction}
The production of lightweight parts for the automotive industry has continuously been gaining importance in recent years for fulfilling the more and more restrictive requirements of low fuel consumption and improved range in ICE and electric vehicles. In this connection, technologies like injection and compression molding, which allow a high volume production of parts, are extensively used in the industry. Specifically, Resin Transfer Molding (RTM) offers better structural capabilities compared to the former ones by virtue of the better control of the fiber alignment and the use of long fibers \cite{Mazumdar_2002}. One important factor in the RTM process is the temperature control during the process as it directly affects the resin viscosity and highly influences the impregnation and the pressure injection. A poor control causes defects like fiber wash-out, preform misalignment and race-tracking \cite{Brocks_2013}.

The present paper describes the development of a temperature controller for a RTM tool heated by distributed cartridge resistances. In this case, the mold will be used for research activities on the filling and curing processes. In consequence, the normally strict temperature control required in RTM tools is even more important in the case under analysis. Specifically, two conditions have to be fulfilled: firstly, the spatial homogeneity of the temperature inside the mold cavity all along the process; and secondly, the accurate tracking of the temperature levels during the rise-up and maintenance phases of the tests. To do that, the controller has to handle in an efficient manner the difficulties derived from the thermal distribution in the cavity, the presence of time-varying variables and internal perturbations (temperature dependent convection cooling, the exothermic nature of the curing heat) and the limited capabilities of the heaters.

Different types of molds are already widespread in the industry and it is therefore possible to find commercial solutions for thermal control like \cite{Watlow_2017}. These alternatives are general developments and normally rely on standard PID compensators \cite{Akashi_1986}. Given its technical importance, more advanced solutions can also be found in the scientific bibliography. These approaches add the knowledge of the system in the controller itself for improving its performance. This is the case of \cite{Wang_2006} and \cite{Munoz_2012}, which use fuzzy controls in plastic injection molds, and \cite{Deng_2012}, which applies neural networks (NN). In \cite{Bosselmann_2017}, a decentralized PI-controller is combined with a feedforward term calculated with a linearized model in a vulcanization test bench. In the present work, MPC is used. This type of controller includes a validated representation of the system for optimizing the control commands also taking into account the characteristics of the actuators. The use of MPC for controlling the temperature of different types of molds has already been reported in the literature. \cite{Lakshmi_2014} and \cite{Gustafson_1987} used this technology in plastic injection applications. In the first case, MPC and IMC are compared for controlling the barrel of a plastic injection machine. To do that, a simplified mold representation based on different transfer functions depending on the operation point are used. In the second case, three simple linear models describe the system dynamics using data directly extracted from a running machine. The controller proposed in the present paper uses a different approach partly due to the different process technology (RTM). As it has to assure the temperature homogeneity inside the cavity, a discretized model in different geometrical regions is used. This model also considers the couplings between the sensors and the actuators. Apart from that, the model faces the nonlinearities in the system by using a perturbation observer therefore simplifying the reference model, which remains linear and independent from the operation point. 

The present paper extends the results from \cite{Escolano_2017} where the basic MPC controller was described and includes the improvements obtained by adding an observer for increasing the resolution of the temperature estimation in the cavity and including the symmetry effects in the optimization process. In overall, the proposed controller is capable of dealing with complex and realistic industrial geometries by combining ARX models with validated FE simulations. The controllers described in the present paper have been successfully implemented on a real mold.

This paper is structured as follows: system description (section \ref{sect_syst_descrip}), thermal model of the mold and its reduced order representation (section \ref{sect_syst_model}), design of the control architectures (section \ref{sect_control_archi}), experimental analysis (section \ref{sect_expvalid}) and conclusions (section \ref{sect_conclusions}).

\section{System description}
\label{sect_syst_descrip}
\begin{figure}[h]
	\centering
		\includegraphics[width=0.5\columnwidth]{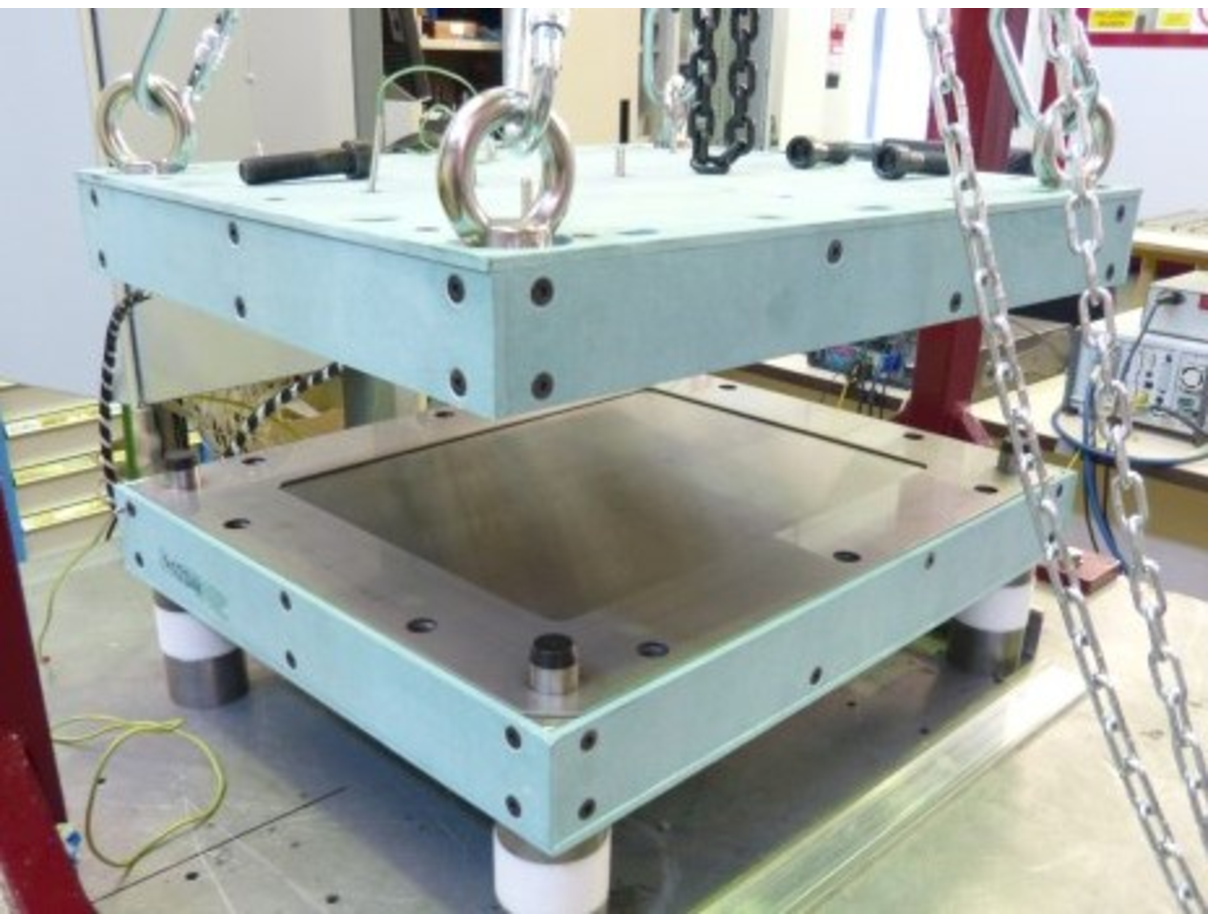}
		\includegraphics[width=0.5\columnwidth]{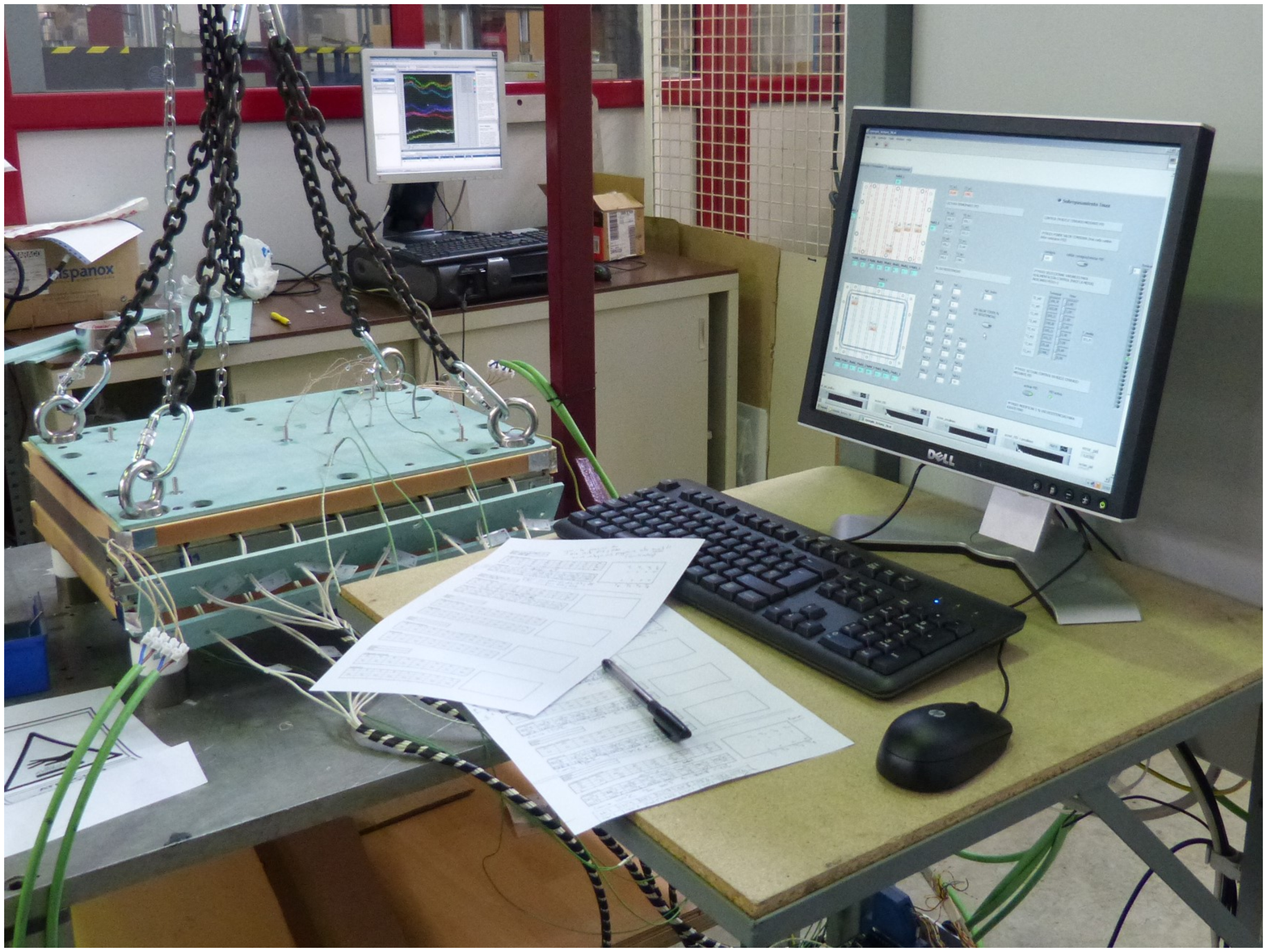}
	\caption{RTM tool and control hardware and software}
	\label{RTM_mold}
\end{figure}

In this section, the main features of the RTM tool and its control hardware are briefly described (Fig. \ref{RTM_mold}).

The RTM tool consists on two steel blocks separated by a spacer which defines the height of the cavity where the reinforcement fabrics are placed and the resin is injected. The cavity is designed to produce composite material panels of 400x300mm with a thickness that can be adjusted using spacers of different height (commonly 2, 3 or 5 mm).

\begin{figure}[H]
	\centering
	\includegraphics[width=\columnwidth]{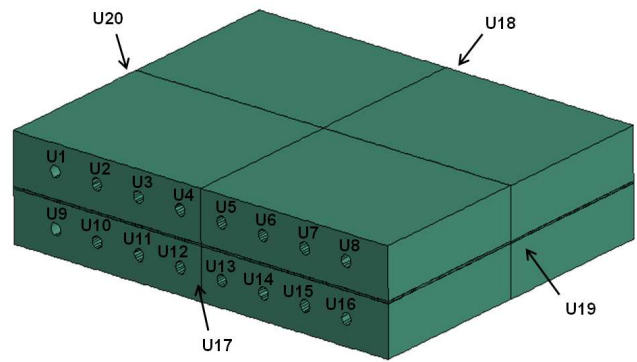}
	\caption{RTM tool sketch. Position and numeration of the heaters.}
	\label{architecture_numbers}
\end{figure}

For heating the tool, the system has 16 internal cartridge resistances (8 in the upper and 8 in the lower part) and 4 lateral heating belts (one per lateral face) as it appears in Fig. \ref{architecture_numbers}. The maximum power that the actuators can supply is 500 W for the cartridge resistances (from $U1$ to $U16$), 750 W for the 2 large heater belts ($U17-U18$) and 550 W for the 2 short heater belts ($U19-U20$). The power of the 20 heaters can be independently commanded by PWM signals.

\begin{figure}[h]
	\centering
		\includegraphics[width=0.5\columnwidth]{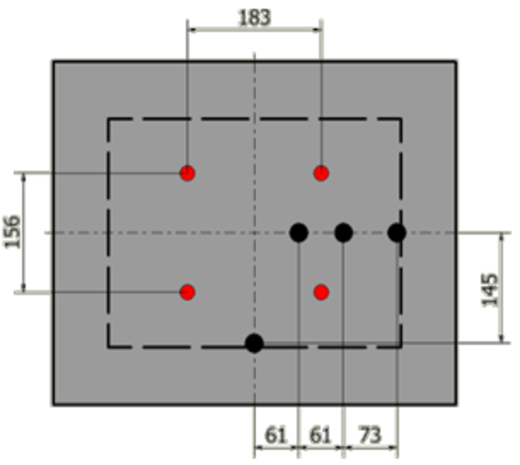}
		\includegraphics[width=0.5\columnwidth]{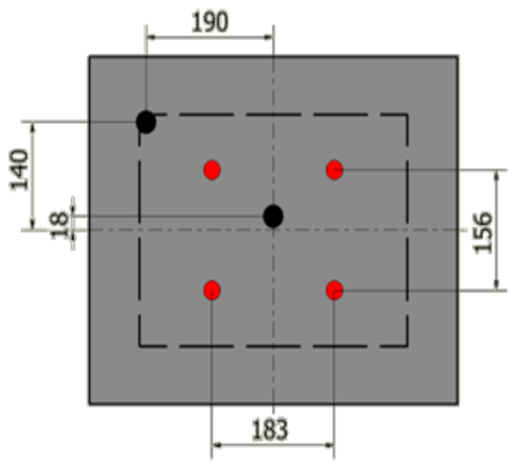}
	\caption{Sketch for the position of the permanent (black dots) and auxiliary (red dots) thermocouples on the surfaces of the upper cavity (left) and lower cavity (right). Dashed line represents the composite material panel.}
	\label{thermo_position}
\end{figure}

6 permanent thermocouples (also called control sensors in this paper) are embedded in the mold cavity (4 in the upper and 2 in the lower cavity) in order to measure the temperature, which is afterwards used by the controller for calculating the required actuation in the heaters. The homogeneity of the temperature distribution in the cavity, and therefore the control performance, is evaluated with 8 additional (also called auxiliary) thermocouples. Figure \ref{thermo_position} shows the position of the permanent and auxiliary thermocouples on the cavity surfaces.

In addition, the heat losses are reduced by insulation panels. They have a thickness of 6 mm in the upper and lower faces and 7 mm in the lateral ones.

The MPC controller is implemented in LabVIEW using control algorithms previously developed in MATLAB/Simulink.

The configuration of resistances, sensors and insulation has been selected based on prior thermal analysis combining simulation and experimental validations.

\section{System modeling}
\label{sect_syst_model}
The detailed FE model of the thermal system is described in section \ref{subsect_continuous} and its reduced order representation appears in section \ref{subsect_ROM}.

\subsection{Continuous model}
\label{subsect_continuous}
The mathematical description of the mold is formulated by applying the principle of energy conservation. The resulting model has the form of a PDE model and its numerical solution is addressed by adopting a Finite Element Model (FEM) discretization method.

\begin{figure}[H]
	\centering
	\includegraphics[width=\columnwidth]{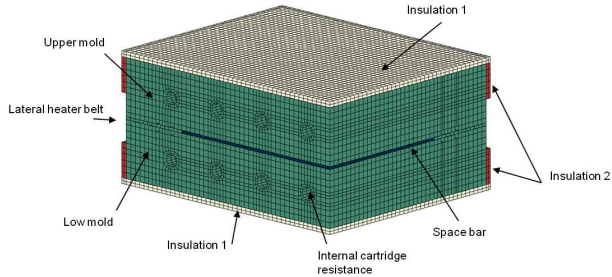} 
	\caption{FEM discretization of the geometry in a quarter of the mold.}
	\label{FEM_discretization}
\end{figure}

The thermal model is discretized in 76952 nodes and the considered heat transfer mechanisms are conduction and convection. The later is represented by effective convection coefficients in the external surfaces of the mold. Figure \ref{FEM_discretization} shows the FE discretization of the geometry in a quarter of the mold.

The material of the steel mold is assumed to be homogeneous and isotropic. Its thermal properties are contained in table \ref{parameter_table}.

\begin{table}[H]
\caption{Thermal properties of steel}
\label{parameter_table}
\centering
\begin{tabular}{|cl|}
	\hline
	Property&Value\\
	\hline
	Density, $\rho(kg/m^{3})$ & 7850\\
	Specific heat, c $(J/kgK)$ & 520\\
	Thermal conductivity, $(W/mK)$ & 33-35.5\\
	\hline
\end{tabular}
\end{table}

In order to obtain the real values of the convection and the insulation conduction coefficients, an extensive set of experimental tests and simulations has been carried out. This identification and validation process considers the thermal behavior of the empty mold in stationary state at different temperature levels.

The fitted thermal conductivity coefficients for the insulation panel 1 and 2 are, respectively, 0.53 $W/mK$ and 0.26 $W/mK$.

The evolution of the convection coefficients of the upper and lower faces has been fitted with Eq.(\ref{conv_ec1}). The coefficient of the lateral faces has been adjusted with Eq.(\ref{conv_ec2}). The identified values of the coefficients (a, b and c) are shown in the table \ref{convenction_coeff}. According to them, the range of values of the convection coefficients for the different surfaces are: 6-18 $W/m^{2}K$ on the upper face, 1-8 $W/m^{2}K$ on the lower face and 2-12 $W/m^{2}K$ on the lateral faces. Out of the analyzed temperature range (below 25\degree C and above 180\degree C), the coefficients are assumed to be constant.

\begin{equation}
h=a\left(\Delta T-b\right)^{c}
\label{conv_ec1}
\end{equation}

\begin{equation}
h=a(b - e^{-c\Delta T})
\label{conv_ec2}
\end{equation}

\begin{table}[H]
	\caption{Values of the convection coefficients in the mold surfaces ($h$ in $W/m^{2}K$ and $\Delta T$ in $K$).}
	\label{convenction_coeff}
	\centering
	\begin{tabular}{|cccc|}
		\hline
		&a&b&c\\
		\hline
		Upper & 4.120 &	23.567&	0.317\\
		Lower & 0.942&	22.937&	0.533\\
		Lateral & 20.160&	0.395&	0.041\\
		\hline
	\end{tabular}
\end{table}

Additionally, the thermal model takes into account the heat released by the exothermic curing process of the composite material panel. The curing kinetic of the resin has been modeled by means of the Kamal-Sourour equation.

The whole FEM model has been implemented in the software Abaqus/Standard.

\subsection{Reduced Order Model (ROM)}
\label{subsect_ROM}
The previous FE representation is complex and computationally expensive to be run in real time as it is required by the Model Predictive Controller. Therefore, a Reduced Order Model based on ARX is built using the data obtained from FE transient simulations. For these virtual tests, the temperature of the mold cavity measured by the 6 permanent sensors ($y$, outputs) is registered when heating power is supplied by the 20 actuators ($u$, inputs). Equation (\ref{ARX_eq}) shows the definition of the ARX model.

\begin{equation}
\begin{array}{lcl}
y_{t+1}&=& \sum_{i=0}^{r-1}a_i y_{t-i} + \sum_{i=0}^{s}b_i u_{t-i}
\end{array}
\label{ARX_eq}
\end{equation}

Where $a$ and $b$ are the parameters of the model, and $r$ and $s$ are the regression order for the outputs and inputs, respectively. 

\begin{figure}[H]
	\centering
	\includegraphics[width=\columnwidth]{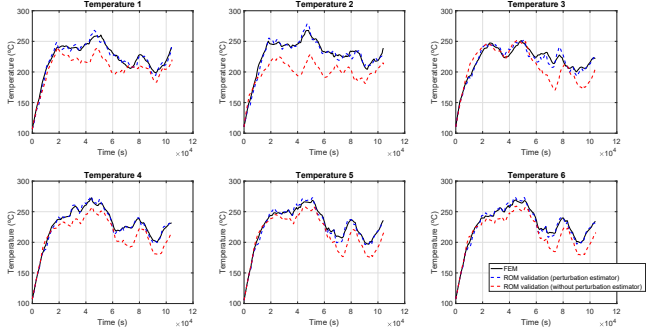}
	\caption{ROM validation with respect to Finite Element Model, before and after the implementation of Kalman filter perturbation estimator}
	\label{estimator_comparison}
\end{figure}

A limitation of the ARX description is its linear nature, which cannot handle the time-varying nature of the convection parameters. In consequence, an initial ROM is obtained using FE results with constant convection coefficient: 15 $W/m^{2}K$ is selected as a mean reference value for natural convection. Significant temperature differences are obtained when this initial ROM is compared with the validated FE model, which includes variable convection coefficients according to Eq.(\ref{conv_ec1}) and Eq.(\ref{conv_ec2}). The identification errors may reach up to 13\% (Fig. \ref{estimator_comparison}). In order to predict and compensate these deviations, a perturbation observer based on a Kalman filter is implemented. 

The ARX model in Eq.(\ref{ARX_eq}) can be written as the state space representation in Eq.(\ref{ROM_ss_eq}):

\begin{equation}
\begin{array}{lcl}
X_{t+1}&=&A X_{t} + B U_{t}\\
Y_{t}&=&C X_{t}
\end{array}
\label{ROM_ss_eq}
\end{equation}

Where,

\begin{equation*}
\begin{array}{lcl}
X_{t}&=&[Y_{t}; Y_{t-1}; Y_{t-2}; ... ; U_{t-1}; U_{t-2}; U_{t-3}; ...]
\end{array}
\end{equation*}

This model assumes constant convection coefficient (15 $W/m^{2}K$). If convection coefficients are not constant, the effect can be described as perturbations in the Eq.(\ref{ROM_ss_eq_pert}).

\begin{equation}
\begin{array}{lcl}
X_{t+1}&=&A X_{t} + B U_{t} + B_p P_{t}\\
Y_{t}&=&C X_{t}
\end{array}
\label{ROM_ss_eq_pert}
\end{equation}

Where $P_{t}$ represents the convection heat differences between 15 $W/m^{2}K$ constant coefficient and variable coefficients according to Eq.(\ref{conv_ec1}) and Eq.(\ref{conv_ec2}).
The state-space representation is then updated by introducing the perturbations $P_t$ as additional state variables:

\begin{equation*}
\begin{array}{lcl}
X_{m,t}&=&[Y_{t}; Y_{t-1}; Y_{t-2}; ... ; U_{t-1}; U_{t-2}; U_{t-3}; ... ; P_t]
\end{array}
\end{equation*}

The final state-space representation is:

\begin{equation}
\begin{array}{lcl}
X_{m,t+1}&=&A_{m}X_{m,t}+B_{m}U_{t}\\
Y_{m,t}&=&C_{m}X_{m,t}
\end{array}
\label{ROM_ss_eq_ext}
\end{equation}

Where,
\begin{equation*}
A_{m}=
\begin{pmatrix}
A & B_p\\
0 & 1
\end{pmatrix}
\end{equation*}

\begin{equation*}
B_{m}=
\begin{pmatrix}
B\\
0
\end{pmatrix}
\end{equation*}

\begin{equation*}
C_{m}=
\begin{pmatrix}
C & 0
\end{pmatrix}
\end{equation*}

The Kalman filter algorithm is applied to the Eq.(\ref{ROM_ss_eq_ext}) state-space, described in Eq.(\ref{Kalman1}-\ref{Kalman5}).

\begin{enumerate}
\item Prediction:
\begin{equation}
\begin{array}{lcl}
\tilde{X}_{m,t}&=&A_{m}\hat{X}_{m,t-1}+B_{m}U_t
\end{array}
\label{Kalman1}
\end{equation}

\begin{equation}
\begin{array}{lcl}
\tilde{P}_{k,t}&=&A_{m}P_{k,t-1}A_{m}^t+C_{q}
\end{array}
\label{Kalman2}
\end{equation}
\item Update:
\begin{equation}
\begin{array}{lcl}
K_{t}&=&\tilde{P}_{k,t}C_{m}^t(C_{m}\tilde{P}_{k,t}C_{m}^t+C_{s})^{-1}
\end{array}
\label{Kalman3}
\end{equation}

\begin{equation}
\begin{array}{lcl}
\hat{X}_{m,t}&=&\tilde{X}_{m,t}+K_t(z_t - C_m\tilde{X}_{m,t})
\end{array}
\label{Kalman4}
\end{equation}

\begin{equation}
\begin{array}{lcl}
P_{k,t}&=&(I-K_tC_m)\tilde{P}_{k,t}
\end{array}
\label{Kalman5}
\end{equation}

\end{enumerate}
Where $C_q$ y $C_s$ are the uncertainties of the estimation model and the sensors respectively; $P$ is the covariance of the estimated state; $K$ is the gain of the Kalman filter; and $z$ is the temperature measurement of the sensors.

After compensating the perturbations, temperature errors lower than 6\% are achieved in the ROM validation. Figure \ref{estimator_comparison} shows the improvement in the ROM validation once perturbations are compensated.

\section{Control architectures}
\label{sect_control_archi}
In this section, three different architectures are described for the temperature control of the RTM tool: 
\begin{enumerate}
\item Standard MPC controller in section \ref{subsect_stdMPC}.
\item Extended domain MPC controller: an augmented observer has been implemented for estimating the temperature of additional points of the mold cavity, in section \ref{subsect_improvedMPC}.
\item Symmetric actuation MPC controller: in addition to the previous approaches, symmetry condition in the power commands is included, in section \ref{subsect_improvedMPC_sym}. 
\end{enumerate}

\subsection{Standard MPC controller}
\label{subsect_stdMPC}
\begin{figure}[H]
	\centering
	\includegraphics[width=\columnwidth]{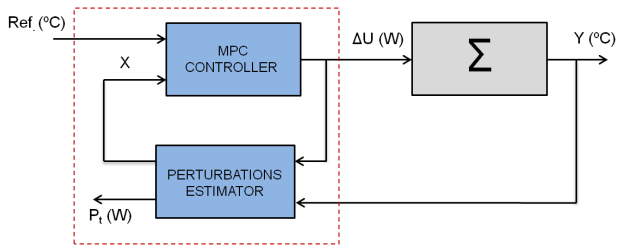} 
	\caption{Control architecture: MPC controller and perturbation estimator.}
	\label{architecture_std}
\end{figure}

A MPC controller is designed for commanding the optimal power to the heaters in order to minimize the temperature differences into the mold cavity. To do that, the ROM and the perturbation observer are used. The input signal of the MPC controller is the state estimation of the system after compensating the perturbations $P_t$ by the Kalman filter estimator described in the previous section. The control architecture is shown in Fig. \ref{architecture_std}.

The MPC controller includes two algorithms: firstly, the estimation of the system evolution predicted by the ROM up to a time horizon defined by $N_p$; and secondly, the calculation of the heating powers by the Hildreth optimization algorithm.

The state-space representation shown in Eq.(\ref{ROM_ss_eq_pert}) has been adapted for its implementation: the inputs are power increments ($\Delta U_t=U_t-U_{t-1}$) and the outputs are integrated inside the state vector itself, according to the Eq.(\ref{MPC_sv_ext}).

\begin{equation}
\begin{array}{lcl}
X_{e,t}&=&[\Delta X_{m,t};Y_{m,t}]
\end{array}
\label{MPC_sv_ext}
\end{equation}
Where $\Delta X_{m,t}= X_{m,t}-X_{m,t-1}$  

Then an extended state-space representation is obtained:

\begin{equation}
\begin{array}{lcl}
X_{e,t+1}&=&A_{e}X_{e,t}+B_{e}\Delta U_{t}\\
Y_{e,t}&=&C_{e}X_{e,t}
\end{array}
\label{MPC_ss_ext}
\end{equation}

Where,
\begin{equation*}
A_{e}=
\begin{pmatrix}
A_{m}& 0\\
C_{m}A_{m} &1
\end{pmatrix}
\end{equation*}

\begin{equation*}
B_{e}=
\begin{pmatrix}
B_{m}\\
C_{m}B_{m}
\end{pmatrix}
\end{equation*}

\begin{equation*}
C_{e}=
\begin{pmatrix}
0 & 1
\end{pmatrix}
\end{equation*}

For the construction of the MPC state-space, the evolution of the following states until the temporal horizon $N_p$ is considered according to the Eq.(\ref{MPC_eq}).

\begin{equation}
\begin{array}{lcl}
Y&=&FX+G\Delta U
\end{array}
\label{MPC_eq}
\end{equation}

Where,

\begin{equation*}
\begin{array}{lcl}
X&=&X_{e,t}
\end{array}
\end{equation*}
\begin{equation*}
\begin{array}{lcl}
Y&=&[Y_{e,t+1}; Y_{e,t+2}; Y_{e,t+3}; ... ; Y_{e,t+N_p}]
\end{array}
\end{equation*}
\begin{equation*}
\begin{array}{lcl}
\Delta U&=&[\Delta U_{t}; \Delta U_{t+1}; \Delta U_{t+2}; ... ; \Delta U_{t+N_p-1}]
\end{array}
\end{equation*}
\begin{equation*}
F=
\begin{pmatrix}
C_{e}A_{e}\\
C_{e}A_{e}^2\\
C_{e}A_{e}^3\\
\vdots\\
C_{e}A_{e}^{N_p}
\end{pmatrix}
\end{equation*}

\begin{equation*}
G=
\begin{pmatrix}
C_{e}B_{e} & 0 & 0 & \cdots & 0\\
C_{e}A_{e}B_{e} & C_{e}B_{e} & 0 & \cdots & 0\\
C_{e}A_{e}^2B_{e} & C_{e}A_{e}B_{e} & C_{e}B_{e} & \cdots & 0\\
\vdots& \vdots& \vdots & \cdots & \vdots \\
C_{e}A_{e}^{N_p-1}B_{e} & C_{e}A_{e}^{N_p-2}B_{e} & C_{e}A_{e}^{N_p-3}B_{e} & \cdots & C_{e}B_{e}
\end{pmatrix}
\end{equation*}

The optimization function $J$ depends on two terms according to Eq.(\ref{MPC_optim}): quadratic error between the reference and the measured temperature at the control sensors, and quadratic term for the power consumption, weighted by $Q$ and $R$ matrices respectively.

\begin{equation}
\begin{array}{lcl}
J&=&(Ref-Y)^T Q (Ref-Y) + \Delta U^T R \Delta U
\end{array}
\label{MPC_optim}
\end{equation}

Where $Ref$ is the vector of reference temperatures until temporal horizon $N_p$.

The Hildreth method is an analytical approach for solving the constrained quadratic optimization problem, based on the resolution of Eq.(\ref{MPC_optim_cond}):

\begin{equation}
\begin{array}{lcl}
\frac{\partial J}{\partial \Delta U} = 0
\end{array}
\label{MPC_optim_cond}
\end{equation}

Taking Eq.(\ref{MPC_optim_cond}), and adapting Eq.(\ref{MPC_eq}) and Eq.(\ref{MPC_optim}), the value of the optimal power increments can be obtained by means of Eq.(\ref{MPC_optimalU}).

\begin{equation}
\begin{array}{lcl}
\Delta U&=&(G^T G + R)^{-1} G^T (Ref - FX)
\end{array}
\label{MPC_optimalU}
\end{equation}

During the optimization, the control commands cannot exceed the maximum values admitted by the resistances and must be higher than zero. In case that the optimization variables do not fulfill the constraints, the algorithm recalculates the commands by the iterative procedure described in \cite{Wang_2009}.

\subsection{Extended domain MPC controller}
\label{subsect_improvedMPC}
The MPC controller described in the previous section only considers the temperature of the control sensors. However, no information is known from the rest of the mold cavity. In order to improve the temperature homogeneity in the whole domain, an augmented Kalman filter estimator is added to the algorithm for estimating the temperature of some cavity points. It includes 8 more points, which will be called virtual nodes hereinafter. The locations of these points correspond with the ones measured by the auxiliary thermocouples (Fig. \ref{thermo_position}).

In this way, the MPC controller calculates the optimal heating powers based, not only on the temperature of the 6 control sensors, but also on the temperature estimation of these 8 virtual nodes ($\hat{T}_{nodes}$). The optimization function $J$ is augmented according to Eq.(\ref{MPC_ext_optim}).

\begin{equation}
\begin{array}{lcl}
J = (Ref-Y)^T Q (Ref-Y) + ... \\
+ (Ref-\hat{T}_{nodes})^T Q (Ref-\hat{T}_{nodes}) + \Delta U^T R \Delta U
\end{array}
\label{MPC_ext_optim}
\end{equation}

\subsection{Symmetric actuation MPC controller}
\label{subsect_improvedMPC_sym}
Despite the extended temperature domain approach described in the previous section, the temperature differences of some cavity areas could be still minimized. In the present section, the symmetry of the mold is used for obtaining a higher homogeneity in areas where the control sensors are not present. To do that, the commanded heating power of the resistances at symmetrical positions with respect to the central line of the mold are forced to be the same. The conditions are applied to the power demands as constraints in the optimization method. According to the numeration followed in Fig. \ref{architecture_numbers}, the expressions described in Eq.(\ref{Sym_numeration}) are implemented.
 
 \begin{equation}
 \begin{array}{cccc}
 U1=U8 & U2=U7 & U3=U6 & U4=U5  \\
 U9=U16& U10=U15 &U11=U14 &U12=U13 \\
  & U17=U18 & U19=U20 &
 \end{array}
 \label{Sym_numeration}
 \end{equation}

\section{Experimental analysis}
\label{sect_expvalid}
The previous MPC controllers have been implemented on a real mold and the obtained results are described in this section. In order to compare the performance, two different conditions are used:
\begin{enumerate}
\item Temperature control of an empty mold (section \ref{subsect_expvalid_empty}).
\item Temperature control in molding conditions (section \ref{subsect_expvalid_fabric}).
\end{enumerate}


For this analysis, the controller updates the power commands each 200s and $N_p$ is fixed to 6. This means that the estimation horizon for the MPC is 1200s. Regarding the optimization function, the weight of matrix $R$ is two orders of magnitude lower than the weight of matrix $Q$ ($Q=1$, $R=0.01$) as temperature homogeneity is the critical indicator for this application.

The obtained results from the experimental analysis are compared in section \ref{subsect_expvalid_comparative}.

\subsection{Experimental analysis in empty mold}
\label{subsect_expvalid_empty}
The three developed MPC controllers are compared when controlling the temperature of the mold cavity in empty conditions. The test sequence is the following one: firsty, the mold is heated from room temperature (23\degree C) up to 120\degree C at 2\degree C/min rate; at 120\degree, the temperature remains constant until t=10000s; then the mold is heated again up to 180\degree C at 2\degree C/min rate; at 180 \degree C, temperature remains constant until t=20000s.

\subsubsection{Standard MPC controller}
\label{subsubsect_expvalid_empty_stdMPC}
\begin{figure}[h]
	\centering
	\includegraphics[width=0.9\columnwidth]{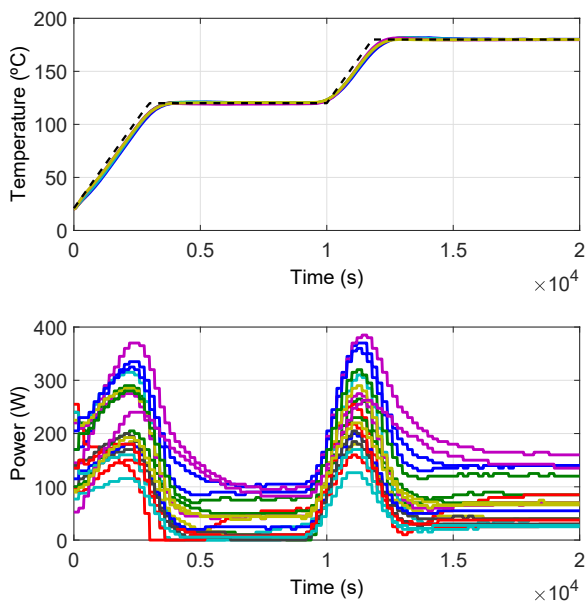}
	\caption{Temperature tracking by the control sensors (upper) and power commands (below) for standard MPC controller.}
	\label{results_std}
\end{figure}

\begin{figure}[h]
	\centering
	\includegraphics[width=0.9\columnwidth]{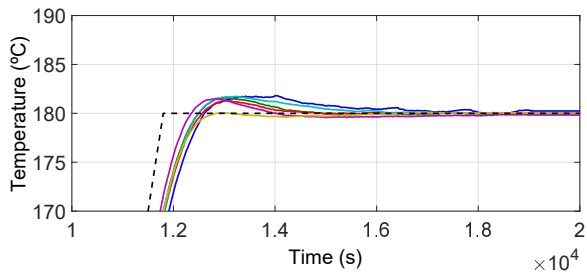}
	\caption{Temperatures measured by the control sensors at 180\degree C reference (standard MPC controller).}
	\label{results_std_sensors}
\end{figure}

\begin{figure}[h]
	\centering
	\includegraphics[width=0.9\columnwidth]{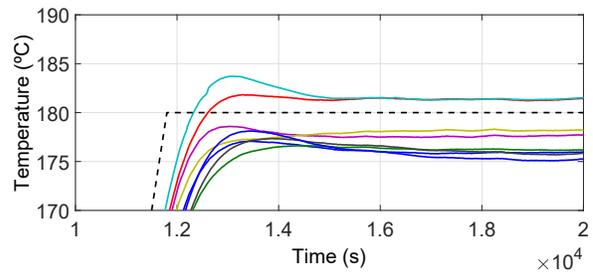}
	\caption{Temperatures measured by the auxiliary thermocouples at 180\degree C reference (standard MPC controller).}
	\label{results_std_thermo}
\end{figure}

Figure \ref{results_std} shows the temperature evolution measured by the 6 control sensors when the heating powers calculated by the standard MPC controller are commanded. In Fig. \ref{results_std_sensors} and Fig. \ref{results_std_thermo}, the temperatures measured by the control sensors and by the auxiliary thermocouples at 180\degree C reference are displayed respectively. 

\subsubsection{Extended Domain MPC controller}
\label{subsubsect_expvalid_empty_improvedMPC}

\begin{figure}[h]
	\centering
	\includegraphics[width=0.9\columnwidth]{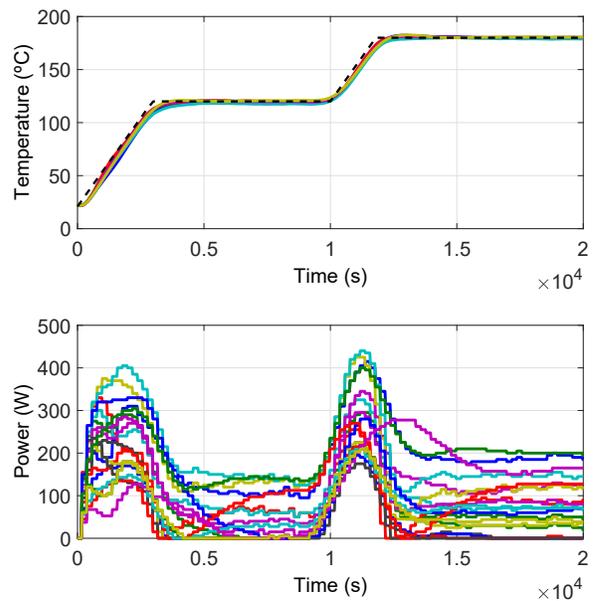}
	\caption{Temperature tracking by the control sensors (upper) and power commands (below) for extended domain MPC controller.}
	\label{results_improved}
\end{figure}

\begin{figure}[h]
	\centering
	\includegraphics[width=0.9\columnwidth]{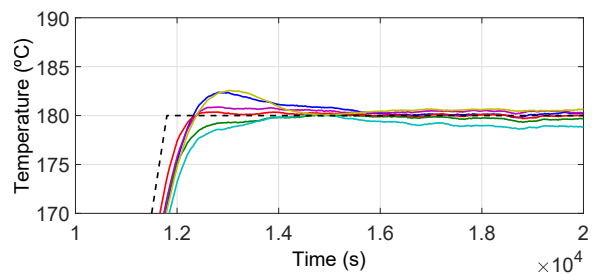}
	\caption{Temperatures measured by the control sensors at 180\degree C reference (Extended domain MPC controller).}
	\label{results_improved_sensors}
\end{figure}

\begin{figure}[h]
	\centering
	\includegraphics[width=0.9\columnwidth]{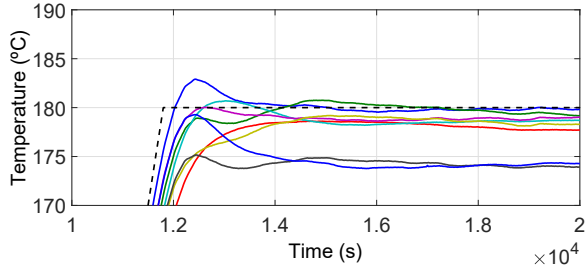}
	\caption{Temperatures measured by the auxiliary thermocouples at 180\degree C reference (Extended domain MPC controller).}
	\label{results_improved_thermo}
\end{figure}

Figure \ref{results_improved} shows the temperature evolution measured by the 6 control sensors when the heating powers calculated by the extended domain MPC controller are commanded. In the Fig. \ref{results_improved_sensors} and Fig. \ref{results_improved_thermo}, the temperatures measured by the control sensors and by the auxiliary thermocouples at 180\degree C reference are displayed respectively. 


\subsubsection{Symmetric actuation MPC Controller}
\label{subsubsect_expvalid_empty_improvedMPC_sym}

\begin{figure}[h]
	\centering
	\includegraphics[width=0.9\columnwidth]{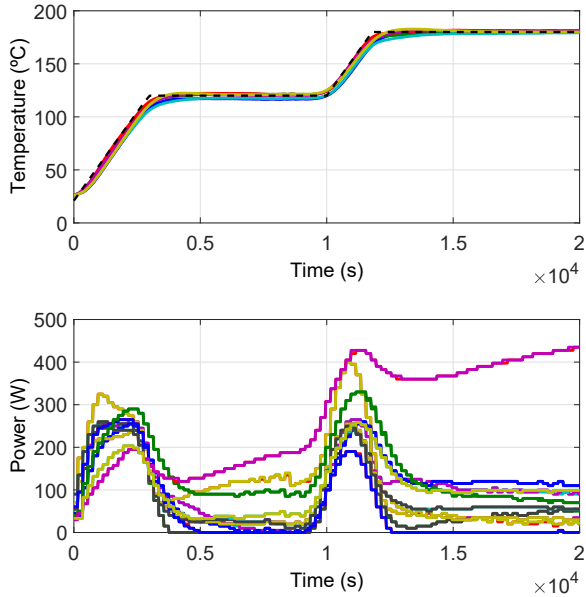}
	\caption{Temperature tracking by the control sensors (upper) and power commands (below) for symmetric actuation MPC controller}
	\label{results_improved_sym}
\end{figure}

\begin{figure}[h]
	\centering
	\includegraphics[width=0.9\columnwidth]{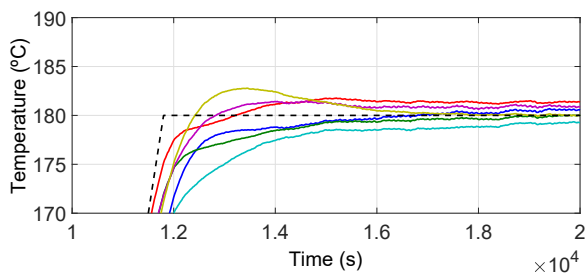}
	\caption{Temperatures measured by the control sensors at 180\degree C reference (Symmetric actuation MPC controller).}
	\label{results_improved_sym_sensors}
\end{figure}

\begin{figure}[h]
	\centering
	\includegraphics[width=0.9\columnwidth]{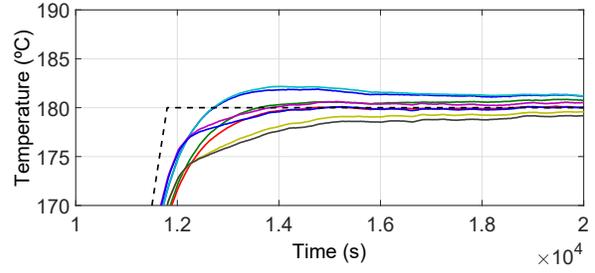}
	\caption{Temperatures measured by the auxiliary thermocouples at 180\degree C reference (Symmetric actuation MPC controller).}
	\label{results_improved_sym_thermo}
\end{figure}

Figure \ref{results_improved_sym} shows the temperature evolution measured by the 6 control sensors when the heating powers calculated by the symmetric actuation MPC controller are commanded. In Fig. \ref{results_improved_sym_sensors} and Fig. \ref{results_improved_sym_thermo}, the temperatures measured by the control sensors and by the auxiliary thermocouples at 180\degree C reference are displayed respectively.  


\subsection{Experimental analysis in molding conditions}
\label{subsect_expvalid_fabric}

The molding of a composite material panel has been carried out by applying the symmetric actuation MPC controller, which is the best approach for minimizing temperature differences in the entire mold cavity. Due to the curing process, an internal heat source from the chemical reactions affects to the temperature controller as a perturbation.

The evolution of the reference temperature for this test is the following one: firstly, the mold is heated from room temperature (23\degree C) up to the injection temperature (120\degree C) at 2\degree C/min rate; once the technician detects a stable stationary level, the resin is injected; the mold is heated up to curing temperature (185\degree C) at 2\degree C/min rate; finally the curing of the resin is ensured maintaining at 185\degree C during two hours.

\begin{figure}[h]
	\centering
	\includegraphics[width=0.9\columnwidth]{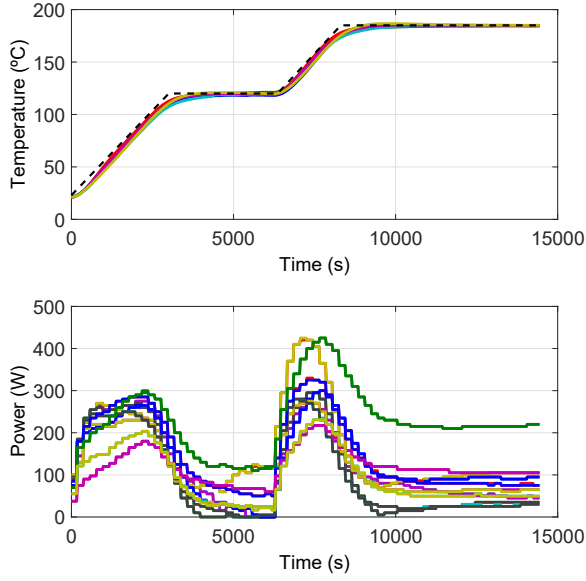}
	\caption{Temperature tracking by the control sensors (upper) and power commands (below) for composite material panel forming.}
	\label{results_fabric}
\end{figure}

\begin{figure}[h]
	\centering
	\includegraphics[width=0.9\columnwidth]{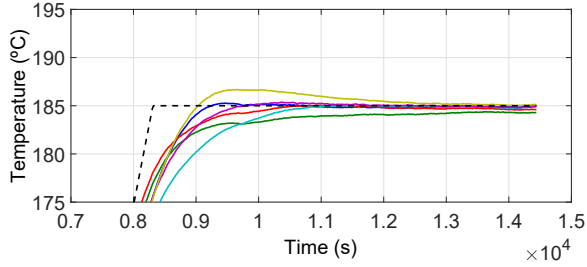}
	\caption{Temperatures measured by the control sensors at 185\degree C reference (composite material panel forming).}
	\label{results_fabric_sensors}
\end{figure}

\begin{figure}[h]
	\centering
	\includegraphics[width=0.9\columnwidth]{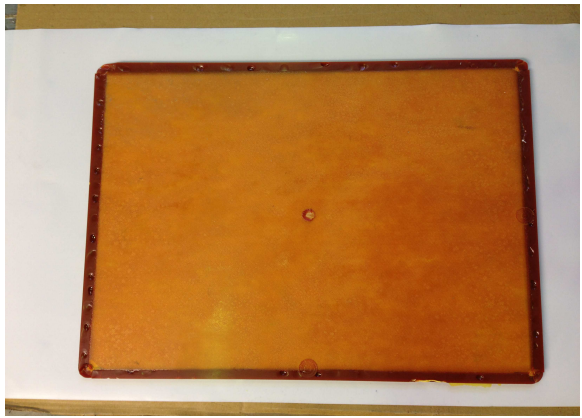}
	\caption{Composite material panel formed by applying MPC controller.}
	\label{results_composite}
\end{figure}

Figure \ref{results_fabric} shows the temperature evolution measured by the 6 control sensors. Figure \ref{results_fabric_sensors} shows the temperatures at 185\degree C reference. In this case, it is not possible to measure the temperature of other cavity points by the auxiliary thermocouples because of the resin injection. Figure \ref{results_composite} shows the formed composite material panel, where no defects or visible variations in superficial appearance are detected.

\subsection{Comparative of results and discussion}
\label{subsect_expvalid_comparative}

The objective of the controller is the tracking of a reference temperature homogeneously in the entire mold cavity. In order to analyze and compare the experimental results, the temperature domain homogeneity and the tracking error are treated as independent indicators. Additionally, these indicators are obtained for the transient state (during the curing process) and for the stationary state (2 hours after the curing temperature is demanded, $t_f$). 

\begin{enumerate}
	\item Stationary indicator of temperature homogeneity ($RMSE_{avg, stat}$): Root mean squared error of temperature by permanent and auxiliary thermocouples ($T_i$) with respect to their average value ($T_{avg}$) at $t_f$, obtained by Eq.(\ref{RMSE_eq1}).
	
	\begin{equation}
	\begin{array}{lcl}
	RMSE_{avg, stat} = \sqrt{ \frac{1}{n} \sum_{i=1}^{n} (T_{i,t_f}-T_{avg,t_f})^2}
	\end{array}
	\label{RMSE_eq1}
	\end{equation}
	
	\item Stationary indicator of the reference temperature tracking ($RMSE_{ref, stat}$): Root mean squared error of temperature by permanent and auxiliary thermocouples ($T_i$) with respect to reference temperature value ($T_{ref}$) at $t_f$, obtained by Eq.(\ref{RMSE_eq2}).
	
	\begin{equation}
	\begin{array}{lcl}
	RMSE_{ref, stat} = \sqrt{ \frac{1}{n} \sum_{i=1}^{n} (T_{i,t_f}-T_{ref,t_f})^2}
	\end{array}
	\label{RMSE_eq2}
	\end{equation}
	
	\item Global indicator of temperature homogeneity ($RMSE_{avg, global}$): Root mean squared error of temperature by permanent and auxiliary thermocouples ($T_i$) with respect to their mean value ($T_{avg}$) from the beginning of the heating process up to the curing temperature ($t_i$=10000s) to $t_f$, obtained by Eq.(\ref{RMSE_eq3}).
	
	\begin{equation}
	\begin{array}{lcl}
	RMSE_{avg, global} = \sqrt{ \frac{1}{n (t_f-t_i)} \sum_{t=t_i}^{t_f} \sum_{i=1}^{n} (T_{i,t}-T_{avg,t})^2}
	\end{array}
	\label{RMSE_eq3}
	\end{equation}
	
	\item Global indicator of the reference temperature tracking ($RMSE_{ref, global}$): Root mean squared error of temperature by permanent and auxiliary thermocouples with respect to reference temperature value ($T_{ref}$) from the beginning of the heating process up to curing temperature ($t_i$=10000s) to $t_f$, obtained by Eq.(\ref{RMSE_eq4}).
	
	\begin{equation}
	\begin{array}{lcl}
	RMSE_{ref, global} = \sqrt{ \frac{1}{n (t_f-t_i)} \sum_{t=t_i}^{t_f} \sum_{i=1}^{n} (T_{i,t}-T_{ref,t})^2}
	\end{array}
	\label{RMSE_eq4}
	\end{equation}	
\end{enumerate}

The values of the indicators for the applied control algorithms are summarized in the table \ref{indicators} and some points are highlighted.

\begin{table}[H]
	\caption{Indicators for the analyzed control algorithms}
	\label{indicators}
	\centering
	\resizebox{\columnwidth}{!}{
		\begin{tabular}{|ccccc|}
			\hline
			& $RMSE_{avg, stat}$ & $RMSE_{ref, stat}$ & $RMSE_{avg, global}$ & $RMSE_{ref, global}$\\
			\hline
			Empty mold & & & & \\
			\hline
			Standard MPC Controller & 2.09	& 2.44	& 2.27	& 4.73	\\
			Extended domain MPC controller & 2.00 & 2.43	& 2.08	& 3.81	\\
			Symmetric actuation MPC Controller & 0.69	& 0.76	& 1.48	& 2.95\\
			\hline
			Molding of a composite material panel & & & & \\
			\hline
			Symmetric actuation MPC Controller & 0.27	& 0.35	& 1.17	& 3.34\\
			\hline
		\end{tabular}
	}
\end{table}

From the comparison of the three MPC controllers for the empty mold cavity, it is concluded that successive improvements in the standard algorithm make that the tracking of the reference temperature is more accurate. Using the ROM for estimating the temperature of virtual nodes has the effect of reducing the RMSE from 4.73 \degree C to 3.81 \degree C during the curing process, and from 2.44 \degree C to 2.43 \degree C for the stationary state. Additionally, if symmetry is applied in the power commands of the heaters, the RMSE is reduced to 2.95 \degree C during the curing process and to 0.76 \degree C for the stationary state.

The improved algorithms also minimize temperature differences in the entire cavity: the RMSE during the curing process is reduced from 2.27 \degree C to 2.08 \degree C for the extended domain approach and to 1.48 \degree C when symmetry is also considered. This reduction is found in the stationary state too: the RMSE decreases from 2.09 \degree C to 2.00 \degree C and to 0.69 \degree C respectively.

For this mold geometry, applying symmetric power commmands achieves higher temperature homogeneity in the mold cavity than extending the domain by estimating the temperature of virtual nodes. 

For the molding of a composite material panel, the last MPC controller approach is analyzed only considering the RMSE in the control sensors as auxiliary thermocouples can not be disposed. The tracking error is 3.34 \degree C during the curing process and 0.35 \degree C for the stationary state; and deviation in temperature homogeneity is 1.17 \degree C during the curing process and 0.27 \degree C for the stationary state. These results fulfill the maximum allowed deviations for RTM processses, which normally range between 2 and 3ºC.

\section{Conclusions}
\label{sect_conclusions}
The present paper describes three possible approaches for controlling the temperature in a RTM tool by using MPC controllers. The results show how the use of ROM with a perturbation observer permits reducing the complexity of the model. In combination with that, the inclusion of state observers and symmetrical conditions contributes to improve the homogeneous temperature distribution inside the mold cavity. All the developed algorithms have been experimentally analyzed. 

As a summary of the obtained results in empty conditions, the standard MPC controller shows a maximum error between the reference and the measured temperature of 2.27 \degree C. Improving this algorithm by means of the temperature estimation of 8 virtual nodes and applying symmetry conditions to the power demands reduces the deviations to 1.48 \degree C. For this improved approach, 0.69 \degree C RMSE value is obtained once the stationary state is reached. 

This controller has also been analyzed during molding conditions. Tight tracking and temperature homogeneity into the mold have been achieved, resulting in the forming of a panel without visible discontinuities or defects.

In order to improve these results, the redefinition of the number and optimal distribution of sensors and actuators could be addressed in future developments. Adaptive model parameters and failure diagnosis based on detailed models could also be explored as more complex control approaches. In addition, the described methodology in this paper could be applied to other processes, for instance, fluid heating or hybrid systems.

\section*{Acknowledgements}

This work was supported by the Spanish Ministry of Economy, Industry and Competitiveness in the framework of the "National Program for Research Aimed at the Challenges of Society, 2014" and the "INNPACTO National public-private cooperation program, 2012".

\bibliographystyle{asmems4}
\bibliography{biblio}


\end{document}